\documentstyle[twoside,fleqn,espcrc2]{article}

\newcommand{\nn}{\nonumber}

\newcommand{\eqn}[1]{(\ref{#1})}

\newcommand{\pslash}{p\kern-1ex /}
\newcommand{\Dslash}{{\cal D}\kern-1.5ex /}

\newcommand{\bpsi}{\overline{\psi}}

\title{One loop calculation of QCD with domain-wall quarks
\thanks{Talk presented by Y.~Taniguchi}}

\author{
S. Aoki 
and
Y. Taniguchi
\address{Institute of Physics, University of Tsukuba, Ibaraki 305, Japan}}

\begin{document}

\begin{abstract}
We calculate one loop corrections to the
domain-wall quark propagator in QCD.
We show how the wave function
is renormalized in this theory.
Especially we are interested in the behavior of the
massless fermion mode, which exists near the domain wall
at the tree level.
We show that this massless mode is stable against the
quantum correction.
\end{abstract}

\maketitle

\makeatletter
\def\setcaption#1{\def\@captype{#1}}
\makeatother

\section{Introduction}

The theory of domain wall fermion was originally introduced to
treat the chiral gauge theory \cite{Kaplan}.
Apart from this expectation
the domain wall fermion is regarded as a suitable formulation to
treat the vector like massless QCD \cite{Shamir}.
This is because of the following great advantages comparing with
the ordinary Wilson or Kogut-Susskind fermion;
(i)The number of flavors is not fixed.
(ii)The renormalization of mass is multiplicative.
In other words, if a massless mode exists in the tree level it is
stable against the quantum correction.
The first one is almost trivial but the latter is not.
It has been understood from an intuitive discussion and a numerical
simulation \cite{Shamir,Blum-Soni}.
The aim of this paper is to confirm the latter nature,
especially the stability of the massless mode by lattice
perturbation theory.

In this paper we set the lattice spacing $a=1$
and take the $SU(N)$ gauge group with second Casimir $C_2$.
We adopt the domain wall fermion of Shamir type \cite{Shamir}
to describe massless quarks.
This type of domain wall fermion is a variation of the Wilson fermion
with sufficiently many $N_S$ flavors and special form of mass matrix.
The only difference from the Wilson fermion action is
the fermion bilinear term.
In the perturbation theory
the gluon propagator and the gauge interaction terms
are exactly same as those in the ordinary Wilson fermion
perturbation theory \cite{KNS,Karsten-Smit} with $N_S$ flavors.
The peculiar thing to the domain wall fermion is the fermion propagator
which is given as an inverse of the fermion bilinear term,
\begin{eqnarray}
&&
\int_{-\pi}^\pi \frac{d^4 p}{(2\pi)^4}
\bpsi (-p) \Bigl[ \sum_\mu i \gamma_\mu \sin p_\mu
\nn\\
&&
\qquad\qquad
+ W^+ (p) P_+ + W^- (p) P_- \Bigr] \psi (p)
\end{eqnarray}
where 
$P_\pm$ is a projection operator;
$P_\pm = (1 \pm \gamma_5)/2$.
Our mass matrix $W^{\pm} (p)$ is given as
\begin{eqnarray}
&&
W^{+} (p) =
\pmatrix{
-W(p) & 1  &        &       \cr
      &    & \ddots &       \cr
      &    & \ddots & 1     \cr
      &    &        & -W(p) \cr
}
\label{eqn:mass-matrix-p}
\\
&&
W^{-} (p) = (W^{+} (p))^\dagger
\label{eqn:mass-matrix-m}
\\
&&
W(p) = 1-M + \sum_\mu (1-\cos p_\mu).
\end{eqnarray}
In spite of the Dirac mass $M$ in the mass matrix, the above action
describes one massless fermion and $N_S -1$ excited modes
at the momentum region $p_\mu \sim 0$
by virtue of this matrix form.
This means that $M$ is not a physical quark mass
but it is rather a cut-off order unphysical mass like
Wilson mass.
$M$ plays an important role as a parameter of the theory
in order to satisfy $|W(p_\mu \sim 0)|<1$, which is needed for
the existence of a massless mode.

In the following we will see this massless mode is
stable against the quantum correction.

\section{Massless Mode}

Before calculating the one loop correction we show how the massless
mode is guaranteed in the domain wall fermion system.

The easiest way to see the massless mode is to diagonalize the fermion
bilinear term.
The bilinear term can always be diagonalized by rotating the right and
left mode of the Dirac fermion independently
by two unitary matrices $U$ and $V$,
\begin{eqnarray}
\psi_{\rm R} \to U \psi_{\rm R},
\qquad
\psi_{\rm L} \to V \psi_{\rm L}.
\label{eqn:U}
\end{eqnarray}
These two unitary matrices $U$ and $V$ should be chosen to
diagonalize two different hermite mass squared matrices
$W^+ W^-$ and $W^- W^+$.
For example the mass squared eigenvalues are following at $M=0.8$
and $N_S = 20$,
\begin{eqnarray}
( 1.43516, \quad \cdots \quad 0.645049, \quad 10^{-18} ).
\label{eqn:tree-eigenvalues}
\end{eqnarray}
The important point is that there is only one massless mode
and other modes are excited ones with cut-off order mass.
By virtue of this nature the domain wall fermion can be regarded
as one massless Dirac fermion system with extra excited modes
decoupled.

This property of the mass eigenvalues can be understood from the
discussion of the eigenstate equation of the mass squared matrix
$W^- W^+$,
\begin{eqnarray}
( W^- W^+ )_{s t} \xi_t^i = M_i^2 \xi_s^i
\end{eqnarray}
where $\xi_i$ and $M_i^2$ are the i-th eigenstate and eigenvalue.
By substituting mass matrix \eqn{eqn:mass-matrix-p} and
\eqn{eqn:mass-matrix-m}.
we have three types of equations depending on $s$.
\begin{eqnarray}
-W \left( \xi^i_{s+1} + \xi^i_{s-1} \right)
+ \left( 1 + W^2 - M_i^2 \right) \xi^i_s = 0
\label{eqn:intermediate}
\\
-W \xi^i_2 + \left( W^2 - M_i^2 \right) \xi^i_1 = 0
\label{eqn:boundary1}
\\
-W \xi^i_{N_S-1} 
+ \left( 1+W^2 -M_i^2 \right) \xi^i_{N_S} = 0
\label{eqn:boundary2}
\end{eqnarray}
From the equation \eqn{eqn:intermediate} which is valid for
$2 \le s \le N_S-1$
we have two kinds of solutions depending on the eigenvalue.
When the eigenvalue is small $M_i^2 \le (1-W)^2$
we have exponential solution,
\begin{eqnarray}
\xi (s) = A e^{\pm \alpha s}
, \quad
\cosh \alpha = \frac{1+W^2 -M_i^2}{2W}
\end{eqnarray}
When the eigenvalue is in the region
$(1-W)^2 \le M_i^2 \le (1-W)^2$ we have an oscillating solution
\begin{eqnarray}
\xi (s) = A e^{\pm i a s}
, \quad
\cos a = \frac{1+W^2 -M_i^2}{2W}
\end{eqnarray}
And when the eigenvalue is larger;
$M_i^2 \ge (1-W)^2$ we again have an exponential solution.

Besides the general equation \eqn{eqn:intermediate}
the solutions have to satisfy the boundary
conditions given by the equations \eqn{eqn:boundary1} and
\eqn{eqn:boundary2},
\begin{eqnarray}
&& -W \xi_0 + \xi_1 = 0
\label{eqn:boundary3}
\\
&& \xi_{N_S+1} = 0
\label{eqn:boundary4}
\end{eqnarray}
Here we assume that $N_S$ is infinitely large.
First, a single dumping solution can satisfy the first boundary
condition with a dumping factor $W$.
This is nothing but a zero mode solution with $M_i^2 = 0$.
Other dumping solutions cannot satisfy this condition.
On the other hand oscillating ones can always satisfy
\eqn{eqn:boundary3} with suitable $M_i^2$.
Indeed we can solve the conditions in this case.
As a result we have one dumping solution with zero eigenvalue
and oscillating solutions with eigenvalues in the region
$ (1-W)^2 \le M_i^2 \le (1+W)^2 $.
For example the tree level result at $M=0.8$ 
has one small eigenvalue and $N_S-1$ eigenvalues in
the region;
$(1-W)^2 = 0.64 \le M_i^2 \le (1+W)^2 = 1.44$.
This is a good coincidence with the above discussion.

\section{One Loop Calculation}

Now we calculate the one loop correction to the fermion propagator,
which is given by a tadpole diagram with a gluon loop and a half-circle
diagram, in which both gluon and fermion runs.
We are interested in stability of the massless mode
given in the $p_\mu \to 0$ limit.
We only require the leading terms in $p_\mu$ to see the massless mode.

The contribution from the tadpole diagram is 
\begin{eqnarray}
\Sigma^{\rm tadpole}
=
g^2 C_2 T \left( \frac{1}{2} i \pslash + 2 \right) \delta_{s,t}
\label{eqn:tadpole}
\end{eqnarray}
with numerical factor
$T = 0.15461$.
The first term in \eqn{eqn:tadpole} is finite, and the second term
is cut-off order.
We can see that $\Sigma^{\rm tadpole}$ is diagonal in flavor space.

The correction from the half circle diagram
cannot be calculated analytically because of its complicated 
dependence on the flavor index $s, t$, which is brought by
the fermion propagator.
We will evaluate the form of $\Sigma^{\rm half}_{s,t}$
according to \cite{Karsten-Smit} by separating the loop momentum
into two region.
The logarithmically divergent part can be calculated
analytically and other linearly divergent and finite terms are
given by integrating the loop momentum numerically,
where we let $N_S = 20$ for the calculation.
And finally the correction is written in a simple form,
\begin{eqnarray}
\Sigma^{\rm half} =
- \Bigl[
i \pslash \left( I^\pm_{\log} + I^\pm_{\rm finite} \right)
P_\pm
+ M^\pm P_\pm
\Bigr]
\end{eqnarray}
where finite term $I^\pm_{\rm finite}$ and linearly divergent
$M^\pm$ are complicated functions of $s, t$ and $M$.

The logarithmic divergence appears only in the wave function 
and is localized near the boundary
\begin{eqnarray}
I_{\rm log}^+ 
=
\frac{1}{16\pi^2} g^2 C_2 M(2-M) (1-M)^{s+t-2}
\nn\\
\times \left( \ln (\pi^2) +\frac{1}{2} - \ln p^2 \right).
\end{eqnarray}
In the diagonal basis of \eqn{eqn:U}
it is exactly localized in the boundary.
This means that the logarithmic divergence can be renormalized
into the zero mode wave function.
On the other hand, the linear divergence in the mass term should be
treated as an additive quantum correction to the mass matrix.
This is because the tree level mass matrix is cut-off order
and a correction of the same order is always allowed.

In the following we will see whether the massless mode is
preserved against the correction to the mass matrix.
The contribution from the tadpole diagram is to modify the mass
parameter to $M-2g^2 C_2 T$.
Although $M^\pm$ has nontrivial flavor dependence,
it is much smaller ($\sim 18\%$ at most) than that of the tadpole
diagram and it turns out to be a similar form to the tree level mass
matrix; minus values along the diagonal line and plus ones along next to
the diagonal.
%
%
Differently from the tree level mass matrix \eqn{eqn:mass-matrix-p},
it's off-diagonal elements are non-zero.
But they are very small comparing the tadpole contribution
($\sim 7\%$ at most)
and dump exponentially along the off-diagonal line
with dumping factor $0.12$ at $M=0.8$.

By virtue of the similarity to the tree level mass matrix
we may expect that the good nature of the domain wall fermion
is preserved at the one loop level.
Diagonalizing the effective mass matrix numerically we have
following sequence of the mass squared eigenvalues for $M=0.8$,
\begin{eqnarray}
( 2.36322, \quad \cdots \quad 0.306761, \quad 10^{-11} ).
\label{eqn:loop-eigenvalues}
\end{eqnarray}
Here again we have a single zero eigenvalue and $N_S -1$ cut-off
order ones.
Corresponding to them we also have one dumping solution
and $N_S -1$ oscillating ones.

From this peculiar distribution of eigenvalues and eigenstates
we may expect that the same mechanism works as at the tree level
and the above nature is guaranteed.
The one loop effective mass matrix
is well approximated by the same form of the matrix
\eqn{eqn:mass-matrix-p}.
Actually with the approximated value of $W=0.5141$
the numerically solved exact results are well reproduced
as an eigenvalue problem of this simple matrix.
The above $W$ is about
$7\%$ smaller than the dumping factor of the exact zero mode solution,
which should be coincide in our approximation.
This difference may be the effect of the off-diagonal parts
which we neglected.

Our conclusion is that the massless mode is preserved at one loop level
because the mass matrix has approximately the same form as at
the tree level.
This fact also guarantees the other modes have cut-off order mass.
This approximation is supported by the following reasons;
(i) main quantum correction comes from the tadpole diagram,
which is diagonal,
(ii) the off-diagonal parts are small and dump exponentially
off the diagonal line,
(iii) numerically solved mass eigenvalues and eigenstates
are well reproduced by this approximated form of the mass matrix.
For our approximation to work it is essentially important that the
off-diagonal parts dump exponentially.

At last we comment that the finite parts in the wave function subtracted
with the logarithmic divergence can be diagonalized simultaneously.
At $M=0.8$ the Z-factor of the zero mode is given as
$Z=1-0.0216602 g^2 C_2$.

Y.~Tanigchi is a JSPS fellow.

\vspace*{-3mm}

\newcommand{\J}[4]{{\it #1} {\bf #2} (19#3) #4}
\newcommand{\MPL}{Mod.~Phys.~Lett.}
\newcommand{\IJMP}{Int.~J.~Mod.~Phys.}
\newcommand{\NP}{Nucl.~Phys.}
\newcommand{\PL}{Phys.~Lett.}
\newcommand{\PR}{Phys.~Rev.}
\newcommand{\PRL}{Phys.~Rev.~Lett.}
\newcommand{\AP}{Ann.~Phys.}
\newcommand{\CMP}{Commun.~Math.~Phys.}
\newcommand{\PTP}{Prog. Theor. Phys.}
\newcommand{\Suppl}{Prog. Theor. Phys. Suppl.}

\end{document}